\documentclass[10pt]{iopart}
\usepackage{graphicx}
\usepackage{bm}
\usepackage[colorlinks=true,linkcolor=blue,urlcolor=blue,citecolor=blue]{hyperref}
\usepackage{times}
\expandafter\let\csname equation*\endcsname=\relax
\expandafter\let\csname endequation*\endcsname=\relax
\usepackage{amsmath,amssymb}
\usepackage{physics}
\usepackage{cite}
\usepackage[left=1in,right=1in]{geometry}

\begin{document}
\title{Heat Current in Non-Markovian Open Systems}
\author{Ruofan Chen}
\address{College of Physics and Electronic Engineering, and Center for Computational Sciences, Sichuan Normal University, Chengdu 610068, China}
\date{\today}

\begin{abstract}
  We generalize time-evolving matrix product operators method to
  nonequilibrium quantum transport problems. The nonequilibrium
  current is obtained via numerical differentiation of the generating
  functional which is represented as a tensor network. The approach is
  numerically exact and the non-Markovian effects are fully taken into
  account. In the transport process, a part of the heat that flows out
  from a bath flows into the system and other baths, and the rest is
  stored in the system-bath coupling part. We take the spin-boson
  model as a demonstration to show the details of this heat flowing
  and the establishment of a steady current between two baths.
\end{abstract}
\maketitle

\section{Introduction}
Quantum transport at nanoscale has attracted both theoretical and
experimental interests for a long time due to the need in device
application
\cite{meir1992-landauer,brandes2004-charge,wang2007-thermal,wang2013-nonequilibrium,segal2013-qubit,lue2015-current,dorsch2021-heat}.
The quantum effects in low dimensional systems often result in
interesting behaviors. For instance, the rectification of heat
transfer in nanodevices can be realized by system structural asymmetry
\cite{segal2005-spin}, the negative differential thermal conductance
can show up with certain conditions
\cite{segal2006-heat,thingna2012-steady}, the Wiedemann-Franz law may
be violated at low temperature \cite{buccheri2022-violation} and the
current direction can be reversed by strong interparticle interaction
\cite{haeusler2021-interaction}.

Beyond the standard linear response approach, exact analytical results
are rare for nonequilibrium quantum transport problems, thus there is
clearly a considerable need for numerically exact methods. Rigorous
non-Markovian and non-perturbative methods for open quantum system
include quasi-adiabatic propagator path integral (QUAPI)
\cite{makarov1993-tunneling,makri1995-numerical,dattani2012-analytic},
mapping Hamiltonian approach with the density matrix renormalization
group \cite{chin2010-exact,prior2010-efficient}, multi-configuration
time-dependent Hartree method \cite{meyer1990-the,wang2000-basis} and
hierarchical equation of motion
\cite{song2017-hierarchical,shi2018-efficient,wang2020-entangled,ikeda2020-generalization,wang2022-statistical}.

The path integral formalism is one of the most suitable approaches to
study the open systems. For a system linearly coupled to bosonic
baths, the dynamics of the system can be described by the
Feynman-Vernon path integral formalism
\cite{feynman1963-the,weiss1993-quantum,caldeira1983-path}.  In order
to be numerically evaluated, the path integral can be discretized via
the QUAPI method. Based on QUAPI, an iterative tensor multiplication
scheme for open system dynamics is proposed
\cite{makarov1994-path,makri1995-numerical}. The QUAPI method is
numerically exact and fully takes non-Markovian effects into account,
therefore it has been widely used to investigate reduced dynamics of
dissipative systems
\cite{makarov1995-control,golosov1999-efficient,shao2002-iterative,thorwart2005-non,nalbach2009-landau,arceci2017-dissipative}. Such
iterative tensor multiplication scheme is also generalized to
fermionic environments
\cite{segal2010-numerically,agarwalla2017-anderson,chen2019-dissipative,chen2020-landau}
and the corresponding quantum transport problem
\cite{weiss2008-iterative,segal2011-nonequilibrium,bedkihal2012-dynamics,simine2013-path}.

The major disadvantage of the QUAPI iterative tensor multiplication
scheme is that its computational cost scales exponentially with the
correlation length and the size of system Hilbert space. Therefore in
practical calculation one often encounters memory problem. Recently,
it is shown that the QUAPI tensor multiplication scheme can be
represented in terms of matrix product state (MPS) and matrix product
operator (MPO) \cite{strathearn2018-efficient,gribben2022-exact}. Then
the iterative process is amenable to the standard MPS compression
algorithm and the computational cost scaling becomes
polynomially. This algorithm is called time-evolving matrix product
operators (TEMPO) method, which greatly extends the applicability of
QUAPI method. The QUAPI and TEMPO method can be generalized to
evaluate the system correlation function
\cite{shao2002-iterative,joergensen2019-exploiting} and heat
statistics \cite{popovic2021-quantum}, and to the situation where
multiple baths are in the presence \cite{gribben2022-exact}. In
addition, using generating functional technique the bath dynamics can
be related to the system correlation function
\cite{gribben2021-using}.

In this article, we generalize TEMPO method to the nonequilibrium
quantum transport problem. The nonequilibrium current can be obtained
via differentiation of the generating functional. Unlike what is done
in Ref. \cite{gribben2021-using}, where the differentiation of
generating functional is handled analytically and then the bath
dynamics is formulated as an integral over system correlation
function, we directly evaluate the differentiation numerically as in
Refs. \cite{weiss2008-iterative,segal2011-nonequilibrium,bedkihal2012-dynamics,simine2013-path}.
The generating functional is represented as a real-time path integral
in Keldysh formalism and numerically evaluated as a tensor network via
QUAPI and TEMPO method. It should be noted that in such evaluation
process, we find that although the auxiliary variable in generating
functional is a single variable, it needs to be handled consistently
with the system variable in QUAPI scheme, i.e., be treated as a path
segment. The corresponding details refer to
Sec. \ref{sec:quasi-adiabatic-propagator-path-integral-method}.

When the system and bath are coupled, there is an amount of energy
stored in the coupling part. Therefore the heat that flows out from a
bath not only flows into the system and other baths but is also stored
in the system-bath coupling part. This part of energy may be referred
to as interaction energy, which is shown to be important to energy
transport in nonequilibrium open systems
\cite{ludovico2014-dynamical,ludovico2016-periodic,ludovico2016-dynamics,ludovico2018-probing}. We
use the spin-boson model to demonstrate the details of such heat
flowing and the establishment of a steady current between two baths.

This article is organized as follows. The detail of the model is given
in Sec. \ref{sec:model}. The method is presented in
Sec. \ref{sec:method}. The simulations of the spin-boson model with
single and two baths are given in
Sec. \ref{sec:spin-boson-model-with-single-bath} and
\ref{sec:spin-boson-model-with-two-baths}, respectively. Finally
Sec. \ref{sec:conclusions} gives conclusions.

\section{Model}
\label{sec:model}
Here we consider a system plus multiple baths model. The Hamiltonian is
\begin{equation}
  \label{eq:hamiltonian}
  \hat{H}=\hat{H}_d+\sum_{\alpha}(\hat{H}_{\alpha}+\hat{H}_{d\alpha}),
\end{equation}
where $\hat{H}_d$ is the system Hamiltonian and $\hat{H}_{\alpha}$ is the
Hamiltonian of $\alpha$th bath. We consider the Caldeira-Leggett type
environment \cite{caldeira1983-path,caldeira1983-quantum} for which
the bath is described by a collection of harmonic oscillators as
\begin{equation}
  \hat{H}_{\alpha}=\sum_k\omega_{\alpha k}\hat{b}_{\alpha k}^{\dag}\hat{b}_{\alpha k}.
\end{equation}
Here $\hat{b}_{\alpha k}^{\dag}$ ($\hat{b}_{\alpha k}$) creates
(annihilates) a boson of state $k$ in $\alpha$th bath with frequency
$\omega_{\alpha k}$. The coupling between the system and $\alpha$th
bath is given by
\begin{equation}
  \hat{H}_{d\alpha}=\hat{s}_{\alpha}\sum_kV_{\alpha k}(\hat{b}_{\alpha k}+\hat{b}_{\alpha k}^{\dag}),
\end{equation}
where $\hat{s}_{\alpha}$ is a system operator which couples to boson
of state $k$ in $\alpha$th bath with strength $V_{\alpha k}$. Here
$\hat{s}_{\alpha}$ of different $\alpha$ are supposed to commute with
each other. The $\alpha$th bath is characterized by a spectral
function
\begin{equation}
  J_{\alpha}(\omega)=\sum_kV_{\alpha k}^2\delta(\omega-\omega_{\alpha k}).
\end{equation}
We consider Ohmic spectral functions for which
\begin{equation}
  J_{\alpha}(\omega)=\lambda_{\alpha}\omega e^{-\omega/\omega_c},
\end{equation}
where $\lambda_{\alpha}$ is the coupling strength parameter and
$\omega_c$ is the cutoff frequency of the bath. For simplicity, the
cutoff frequency is set to $\omega_c=3.5$ for all baths throughout
this article.

We assume that at initial time $t=0$ the total density matrix is
separable into system and baths for which
\begin{equation}
  \label{eq:initial-rho}
  \hat{\rho}(0)=\hat{\rho}_d(0)\prod_{\alpha}\hat{\rho}_{\alpha},
\end{equation}
where $\hat{\rho}_d(0)$ is the initial system density matrix and the
$\alpha$th bath is in thermal equilibrium for which
$\hat{\rho}_{\alpha}=e^{-\hat{H}_{\alpha}/T_{\alpha}}$. Here
$T_{\alpha}$ is the temperature of $\alpha$th bath. Throughout this
article we set $\hbar=k_B=1$ and use dimensionless quantities.

The heat current flows out from the $\alpha$th bath at time $t$ is
defined as the opposite of the energy change rate for which
\begin{equation}
  I_{\alpha}(t)=-\langle\hat{\dot{H}}_{\alpha}(t)\rangle=
  -i\langle[\hat{H},\hat{H}_{\alpha}(t)]\rangle=
  i\sum_kV_{\alpha k}\omega_{\alpha k}\langle
  \hat{s}_{\alpha}(t)[\hat{b}_{\alpha k}(t)-\hat{b}_{\alpha k}^{\dag}(t)]\rangle.
\end{equation}
The system quantum operator $\hat{s}_{\alpha}$ commutes with the bath
operators $\hat{b}_{\alpha k}$ and $\hat{b}_{\alpha k}^{\dag}$, therefore the
above expression can be written as
\begin{equation}
  \label{eq:current-definition}
  I_{\alpha}(t)=-2\,\mathrm{Im}\sum_kV_{\alpha k}\omega_{\alpha k}
  \langle\hat{s}_{\alpha}(t)\hat{b}_{\alpha k}(t)\rangle.
\end{equation}

The total density matrix $\hat{\rho}(t)$ evolves according to the Von
Neumann equation which gives a formal solution
$\hat{\rho}(t)=e^{-i\hat{H}t}\hat{\rho}(0)e^{i\hat{H}t}$. The reduced
density matrix $\hat{\rho}_d(t)$ is obtained via tracing out the whole
environment from total density matrix
$\hat{\rho}_d(t)=\Tr_{\mathrm{E}}[\hat{\rho}(t)]$.  The knowledge of
reduced density matrix $\hat{\rho}_d(t)$ allows us to calculate the
expectation value of any system operator $\hat{f}(t)$ via
$\langle\hat{f}(t)\rangle=Z_d^{-1}\Tr_d[\hat{\rho}_d(t)\hat{f}]$,
where $Z_d=\Tr_d[\hat{\rho}_d(t)]$ and $\Tr_d$ is the trace over the
system. However, to evaluate the current formula
\eqref{eq:current-definition} we need to calculate the quantity
$\langle\hat{s}_{\alpha}(t)\hat{b}_{\alpha k}(t)\rangle$ which
involves bath operator $\hat{b}_{\alpha k}(t)$. In order to handle
such an operator, we define a reduced density matrix with a source
term as
\begin{equation}
  \hat{\rho}_d^{\xi_{\alpha}}(t)=\Tr_{\mathrm{E}}[\hat{\rho}(t)e^{\xi_{\alpha}\sum_kV_{\alpha k}\omega_{\alpha k}\hat{b}_{\alpha k}}].
\end{equation}
This $\hat{\rho}_d^{\xi_{\alpha}}(t)$ may be referred to as generating
functional in path integral formalism, which shall be discussed in the
following section. Note that $\hat{\rho}_d^{\xi_{\alpha}}(t)$ reduces
to $\hat{\rho}_d(t)$ when $\xi_{\alpha}=0$. Then via differentiation with
respect to $\xi_{\alpha}$ we have
\begin{equation}
  \sum_kV_{\alpha k}\omega_{\alpha k}\langle\hat{s}_{\alpha}(t)\hat{b}_{\alpha k}(t)\rangle=
  Z_d^{-1}\eval{\Tr_d\qty[\fdv{\hat{\rho}_d^{\xi_{\alpha}}(t)}{\xi_{\alpha}}\hat{s}_{\alpha}]}_{\xi_{\alpha}=0}.
\end{equation}
Then the current formula \eqref{eq:current-definition} can be
evaluated numerically with a finite but small $\xi_{\alpha}$ for which
\begin{equation}
  I_{\alpha}(t)=-2\,\Im Z_d^{-1}\Tr_d\qty[\frac{\hat{\rho}_d^{\xi_{\alpha}}(t)-\hat{\rho}_d(t)}{\xi_{\alpha}}\hat{s}_{\alpha}].
\end{equation}

The total heat $Q_{\alpha}(t)$ that flows out from the $\alpha$th bath
at time $t$ is given by a time integral of heat current as
\begin{equation}
  Q_{\alpha}(t)=\int_0^tI_{\alpha}(t')\dd{t'}.
\end{equation}
A part of this heat flows into the system and other baths, and the
rest is stored in the coupling part which may be referred to as
interaction energy. We denote the system energy change at time $t$ as
\begin{equation}
  E(t)=\langle\hat{H}_d(t)\rangle-\langle\hat{H}_d(0)\rangle,
\end{equation}
and the interaction energy between the system and $\alpha$th bath at
time $t$ as
\begin{equation}
  W_{\alpha}(t)=\langle\hat{H}_{d\alpha}(t)\rangle=2\Re\sum_kV_{\alpha k}
  \langle\hat{s}_{\alpha}(t)\hat{b}_{\alpha k}(t)\rangle.
\end{equation}
Since at time $t=0$ the total density matrix \eqref{eq:initial-rho} is
in product state, we have $\langle\hat{H}_{d\alpha}(0)\rangle=0$ and
the quantity $W_{\alpha}(t)$ defined here is also the interaction
energy change
$\langle\hat{H}_{d\alpha}(t)\rangle-\langle\hat{H}_{d\alpha}(0)\rangle$.

Similarly, we can define another reduced density matrix with a source
term as
\begin{equation}
  \hat{\rho}_d^{\chi_{\alpha}}(t)=\Tr_{\mathrm{E}}[\hat{\rho}(t)e^{\chi_{\alpha}\sum_kV_{\alpha k}\hat{b}_{\alpha k}}],
\end{equation}
and a numerical differentiation with a small $\chi_{\alpha}$ yields
\begin{equation}
  W_{\alpha}(t)=2\Re Z_d^{-1}
  \Tr_d\qty[\frac{\hat{\rho}_d^{\chi_{\alpha}}(t)-\hat{\rho}_d(t)}{\chi_{\alpha}}\hat{s}_{\alpha}].
\end{equation}
In this article, we set $\xi_{\alpha}=\chi_{\alpha}=10^{-3}$, then the
error due to numerical differentiation should be at the order of
$O(10^{-3})$ since $[(e^x-1)/x-1]=O(x)$.

The energy conservation is expressed by $Q(t)$, $E(t)$ and $W(t)$ as
\begin{equation}
  \label{eq:heat-conservation}
  \sum_{\alpha}Q_{\alpha}(t)=E(t)+\sum_{\alpha}W_{\alpha}(t),
\end{equation}
or be expressed by the corresponding time derivates as
\begin{equation}
  \label{eq:current-conservation}
  \sum_{\alpha}I_{\alpha}(t)=\dot{E}(t)+\sum_{\alpha}\dot{W}(t).
\end{equation}
The energy conservation is numerically fulfilled for the algorithm
described in this article.

\section{Method}
\label{sec:method}
In this section, we present the basic framework of our method, which
includes the path integral formalism, the discretization (QUAPI
scheme) and the MPS representation (TEMPO scheme) for path integral
formalism.

\subsection{Path Integral Formalism}
\label{sec:path-integral-formalism}
Now we want to express $\hat{\rho}_d(t)$,
$\hat{\rho}_d^{\xi_{\alpha}}(t)$ and $\hat{\rho}_d^{\chi_{\alpha}}(t)$
in path integral representation. In this section, we give a basic
introduction of the path integral formalism, and a more detailed
derivation refers to \ref{app:details-of-path-integral-formalsim}.

\begin{figure}[htbp]
  \centerline{\includegraphics[]{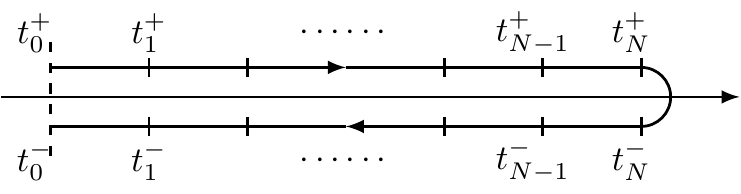}}
  \caption{The Keldysh contour $\mathcal{C}$. The upper is the forward
    branch and the lower is the backward branch.}
  \label{fig:01}
\end{figure}

Splitting the evolution time into $N$ steps that $\delta t=t/N$ with
$N\to\infty$ yields the total density matrix as
\begin{equation}
  \label{eq:rho-split}
  \hat{\rho}(t)=e^{-i\hat{H}\delta t}\cdots e^{-i\hat{H}\delta t}\hat{\rho}(0)e^{i\hat{H}\delta t}\cdots e^{i\hat{H}\delta t}.
\end{equation}
We insert the identity operator into every time step and label the
time steps starting from $\hat{\rho}(0)$ to the leftmost as
$(t_0^+,\ldots,t_N^+)$ and the steps to the rightmost as
$(t_0^-,\ldots,t_N^-)$. Due to the cyclic property
\begin{equation}
  \langle \hat{f}(t)\rangle=\Tr[e^{-i\hat{H}t}\hat{\rho}(0)e^{i\hat{H}t}\hat{f}]=\Tr[e^{i\hat{H}t}\hat{f}e^{-i\hat{H}t}\hat{\rho}(0)],
\end{equation}
we can think the evolution starts from time $0$ then experiences an
evolution $e^{-i\hat{H}t}$, and after a measurement of $f$ returns back to
time $0$ by a backward evolution $e^{i\hat{H}t}$. Therefore the set of time
steps $(t_0^+,\ldots,t_N^+,t_N^-,\ldots,t_0^-)$ forms a closed time
contour $\mathcal{C}$ shown in Fig. \ref{fig:01}, which is usually
referred to as Keldysh contour
\cite{keldysh1965-diagram,lifshitz1981-physical,kamenev2009-keldysh,wang2013-nonequilibrium}. The
path from $t_0^+$ to $t_N^+$ corresponds the forward evolution
operator $e^{-i\hat{H}t}$ and thus we call it forward
branch. Correspondingly, we call the path from $t_N^-$ to $t_0^-$ the
backward branch.

On the contour, we can define free bath Green's functions, which are
useful to describe path integral formalism. If $t'$ succeeds $t''$ on
the contour then we denote it by $t'\succ t''$. With this ordering,
the free bath Green's function on the contour $\mathcal{C}$ is defined
as
\begin{equation}
  G_{\alpha k}(t',t'')=\langle T_{\mathcal{C}}\hat{b}_{\alpha k}(t')\hat{b}^{\dag}_{\alpha k}(t'')\rangle_0,
\end{equation}
where $T_{\mathcal{C}}$ is the contour ordered operator and
$\langle\cdots\rangle_0$ means the expectation value in the free
bath. To be specific,
\begin{equation}
  G_{\alpha k}(t',t'')=\begin{cases}
    \langle \hat{b}_{\alpha k}(t')\hat{b}_{\alpha k}^{\dag}(t'')\rangle_0,&t'\succeq t'';\\
    \langle \hat{b}_{\alpha k}^{\dag}(t'')\hat{b}_{\alpha k}(t')\rangle_0,&t''\succ t';\\
  \end{cases}
\end{equation}
The contour ordered Green's function can be split into four
nonequilibrium Green's function blocks as
\begin{equation}
  G_{\alpha k}(t',t'')=\mqty[
  G_{\alpha k}^{++}(t',t'')&G_{\alpha k}^{+-}(t',t'')\\
  G_{\alpha k}^{-+}(t',t'')&G_{\alpha k}^{--}(t',t'')
  ],
\end{equation}
where in $G^{++}$ ($G^{--}$) both $t',t''$ are on the forward
(backward) branch, and in $G^{+-}$ ($G^{-+}$) $t'$ is on the forward
(backward) branch and $t''$ is on the backward (forward) branch.

The matrix element of the reduced density matrix is
\begin{equation}
  \mel{s'}{\hat{\rho}_d(t)}{s''}=\Tr_{\mathrm{E}}[\mel{s'}{e^{-i\hat{H}t}\hat{\rho}(0)e^{i\hat{H}t}}{s''}].
\end{equation}
Relabeling $s'=s(t_N^+)=s_N^+$ and $s''=s(t_N^-)=s_N^-$, then the
corresponding set $(s_0^+,\ldots,s_N^+,s_N^-,\ldots,s_0^-)$ forms a
path of $s(t)$ on the contour $\mathcal{C}$. Following standard
procedures
\cite{feynman1963-the,grabert1988-quantum,weiss1993-quantum,negele1998-quantum},
the reduced density matrix can be presented as a path integral as
\begin{equation}
  \label{eq:path-integral-rho}
  \rho_d(s_N^{\pm})=\mel{s_N^+}{\hat{\rho}_d(t)}{s_N^-}=(\prod_{\alpha}Z_{\alpha}^{(0)})\int\mathcal{D}[s]K[s]F[s],
\end{equation}
where $Z_{\alpha}^{(0)}$ is the free partition function of the
$\alpha$th bath and $K[s]$ is the propagator of the bare system. Here
the integral over $\mathcal{D}[s]$ indicates the summation over all
the possible path on the contour $\mathcal{C}$ with boundary condition
$s_N^+=s'$ and $s_N^-=s''$.

The term $F[s]$ is the total influence functional which is the
production of influence functional of each bath $F_{\alpha}[s]$ for
which $F[s]=\prod_{\alpha}F_{\alpha}[s]$. The quantity $F_{\alpha}[s]$
is the Feynman-Vernon influence functional of $\alpha$th bath for
which
\begin{equation}
  \label{eq:F}
  F_{\alpha}[s]=e^{-\int_{\mathcal{C}}\dd{t'}\int_{\mathcal{C}}\dd{t''}
    s_{\alpha}(t')\varDelta_{\alpha}(t',t'')s_{\alpha}(t'')},
\end{equation}
where
\begin{equation}
  \label{eq:varDelta-G}
  \varDelta_{\alpha}(t',t'')=\sum_k V_{\alpha k}^2G_{\alpha k}(t',t'').
\end{equation}
The generating functional $\rho_d^{\xi_{\alpha}}(s_N^{\pm})$ can be
expressed as path integral as
\begin{equation}
  \label{eq:generating-functional-xi}
  \rho_d^{\xi_{\alpha}}(s_N^{\pm})=(\prod_{\alpha}Z_{\alpha}^{(0)})\int\mathcal{D}[s]K[s]F[s]X_{\xi_{\alpha}}[s],
\end{equation}
where
\begin{equation}
  \label{eq:X}
  X_{\xi_{\alpha}}[s]=e^{-i\xi_{\alpha}\int_{\mathcal{C}}\dd{t'}s_{\alpha}(t')\varGamma_{\alpha}(t',t_N^-)}
\end{equation}
with
\begin{equation}
  \label{eq:varGamma-G}
  \varGamma_{\alpha}(t',t'')=\sum_kV_{\alpha k}^2\omega_{\alpha k}G_{\alpha k}(t',t'').
\end{equation}
Similarly, the generating functional
$\rho_d^{\chi_{\alpha}}(s_N^{\pm})$ can be expressed as
\begin{equation}
  \label{eq:generating-functional-chi}
  \rho_d^{\chi_{\alpha}}(s_N^{\pm})=(\prod_{\alpha}Z_{\alpha}^{(0)})
  \int\mathcal{D}[s]K[s]F[s]Y_{\chi_{\alpha}}[s],
\end{equation}
where
\begin{equation}
  \label{eq:Y}
  Y_{\chi_{\alpha}}[s]=e^{-i\chi_{\alpha}\int_{\mathcal{C}}\dd{t'}s_{\alpha}(t')\varDelta_{\alpha}(t',t_N^-)}.
\end{equation}
Since $\varDelta_{\alpha}$ and $\varGamma_{\alpha}$ are linear to
$G_{\alpha}$, they can be also split into four blocks as
$\varDelta_{\alpha}^{\pm\pm}$ and $\varGamma_{\alpha}^{\pm\pm}$.

\subsection{Quasi-Adiabatic Propagator Path Integral Method}
\label{sec:quasi-adiabatic-propagator-path-integral-method}
To be evaluated numerically, the path integral formalism
$F_{\alpha}[s]$, $X_{\xi_{\alpha}}[s]$ and $Y_{\chi_{\alpha}}[s]$ must
be discretized. It should be noted that directly employing finite
$\delta t$ may cause a large deviation from the correct result,
therefore an improved discretization scheme, which is called
quasi-adiabatic propagator path integral (QUAPI) method
\cite{makarov1993-tunneling,makri1995-numerical,dattani2012-analytic},
need to be adopted.

The Feynman-Vernon influence functional can be written in a more
specific form as
\cite{feynman1963-the,caldeira1983-path,weiss1993-quantum}
\begin{equation}
  \label{eq:influence-functional}
  F_{\alpha}[s]=e^{-\int_0^t\dd{t'}\int_0^{t'}\dd{t''}[s_{\alpha}^+(t')-s_{\alpha}^-(t')]
    [C_{\alpha}(t'-t'')s_{\alpha}^+(t'')-\bar{C}_{\alpha}(t'-t'')s_{\alpha}^-(t'')]},
\end{equation}
where $C_{\alpha}(t)$ is the autocorrelation function of $\alpha$th
bath for which
\begin{equation}
  \label{eq:autocorrelation}
  C_{\alpha}(t)=\int\dd{\omega}J_{\alpha}(\omega)\qty[\coth(\frac{\omega}{2T_{\alpha}})\cos\omega t-i\sin\omega t],
\end{equation}
and $\bar{C}_{\alpha}(t)$ is the complex conjugate. After the
discretization, the double integral in exponential becomes a double
summation as
\begin{equation}
  \label{eq:influence-functional-discretized}
  F_{\alpha}[s]=e^{-\sum_{j=0}^N\sum_{k=0}^j(s_{\alpha j}^+-s_{\alpha j}^-)
    (\eta_{jk}s_{\alpha k}^+-\bar{\eta}_{jk}s_{\alpha k}^-)},
\end{equation}
where $\eta_{jk}$ is a complex number and $\bar{\eta}_{jk}$ is its
complex conjugate. Here the path $s_{\alpha}^{\pm}(t)$ is split into
intervals of equal duration for which
$s_{\alpha}^{\pm}(t)=s_{\alpha j}^{\pm}$ for
$(j-\frac{1}{2})\delta t<t<(j+\frac{1}{2})\delta t$. In this case we
have for $j\ne k$
\begin{equation}
  \label{eq:eta-jk}
  \eta_{jk}=\int_{(j-\frac{1}{2})\delta t}^{(j+\frac{1}{2})\delta t}\dd{t'}
  \int_{(k-\frac{1}{2})\delta t}^{(k+\frac{1}{2})\delta t}\dd{t''}C_{\alpha}(t'-t''),
\end{equation}
and for $j=k$
\begin{equation}
  \label{eq:eta-jj}
  \eta_{jj}=\int_{(j-\frac{1}{2})\delta t}^{(j+\frac{1}{2})\delta t}\dd{t'}
  \int_{(j-\frac{1}{2})\delta t}^{t'}\dd{t''}C_{\alpha}(t'-t'').
\end{equation}

The source term $X_{\xi_{\alpha}}[s]$ can be written in a specific
form as
\begin{equation}
  X_{\xi_{\alpha}}[s]=e^{-i\xi_{\alpha}\int_0^t\dd{t'}[s_{\alpha}^+(t')\varGamma_{\alpha}^{+-}(t'-t)-
    s_{\alpha}^-(t')\varGamma_{\alpha}^{--}(t'-t)]},
\end{equation}
which should also be discretized via QUAPI scheme. It should be
emphasized that although the variable $\xi_{\alpha}$ is only at a
single time step $t_N^-$, it in fact corresponds to the path segment
$s_N^-$. Therefore we need to replace it by a segment as
\begin{equation}
  \xi_{\alpha}\to\frac{1}{\delta t}\int_{(N-\frac{1}{2})\delta t}^{(N+\frac{1}{2})\delta t}\xi_{\alpha}(t')\dd{t'}
\end{equation}
with $\xi_{\alpha}(t')=\xi_{\alpha}$. Otherwise, we would obtain wrong
results which directly violate energy conservation expression
\eqref{eq:heat-conservation} and \eqref{eq:current-conservation}.
After the discretization we have
\begin{equation}
  \label{eq:X-discretization}
  X_{\xi_{\alpha}}[s]=e^{-i\xi_{\alpha}\sum_{j=0}^N[s^+_{\alpha j}\gamma^{+-}_{jN}-s^-_{\alpha j}\gamma^{--}_{jN}]},
\end{equation}
where for $j\ne N$
\begin{equation}
  \gamma_{jN}^{\pm\pm}=\frac{1}{\delta t}\int_{(N-\frac{1}{2})\delta t}^{(N+\frac{1}{2})\delta t}\dd{t'}
  \int_{(j-\frac{1}{2})\delta t}^{(j+\frac{1}{2})\delta t}\dd{t''}\varGamma^{\pm\pm}_{\alpha}(t''-t'),
\end{equation}
and for $j=N$
\begin{equation}
  \gamma_{NN}^{\pm\pm}=\frac{1}{\delta t}\int_{(N-\frac{1}{2})\delta t}^{(N+\frac{1}{2})\delta t}\dd{t'}
  \int^{t'}_{(N-\frac{1}{2})\delta t}\dd{t''}\varGamma^{\pm\pm}_{\alpha}(t''-t').
\end{equation}

\subsection{Time-Evolving Matrix Product Operators}
\label{sec:time-evolving-matrix-product-operators}

The Feynman-Vernon influence functional
\eqref{eq:influence-functional} is nonlocal in time, and the
nonlocality is characterized by the autocorrelation function
\eqref{eq:autocorrelation}, which corresponds to $\eta_{jk}$ in
formula \eqref{eq:influence-functional-discretized} after the
discretization. For ohmic spectral function, it is shown that
\cite{caldeira1983-path} in the limit $T_{\alpha}\to\infty$ the real
part of the autocorrelation function behaves as the delta function
$\delta(t)$ and its imaginary part behaves as $\delta'(t)$. In this
case, the autocorrelation function is local in time and the dynamics
is just Markovian. At finite temperature, the autocorrelation function
becomes nonlocal but the nonlocality is finite even at low or zero
temperature \cite{makarov1994-path,makri1995-numerical}.

The finite nonlocality means that $\eta_{jk}$ can be truncated when
$\abs{j-k}$ is larger than a certain positive integer $N_s$. Based on
such property, a numerically exact iterative tensor multiplication
scheme based on QUAPI for long time dynamics of quantum dissipative
systems is proposed \cite{makarov1994-path,makri1995-numerical}. The
nonlocal interactions that arise from $\varGamma^{+-}_{\alpha}(t)$ and
$\varGamma^{--}_{\alpha}(t)$ also have finite range, which enables us
to truncate $X_{\xi_{\alpha}}[s]$ and $Y_{\chi_{\alpha}}[s]$ in the
same way.

In the original QUAPI algorithm, the computational cost scales
exponentially with $N_s$. Typical simulations of QUAPI are restricted
to $N_s<20$ \cite{nalbach2011-iterative,thorwart2005-non}, and in fact
when $N_s$ is greater than 10 it already becomes time consuming. In
order to cover the nonlocality, we need $N_s\delta t\ge 4$, and in
such a situation time interval $\delta t$ need often to be not less
than a fairly large value $0.25$.

Recently, it is shown that the QUAPI tensor multiplication scheme can
be represented in terms of matrix product states (MPS) and matrix
product operator (MPO)
\cite{strathearn2018-efficient,gribben2022-exact}. Then the iterative
process is amenable to the standard MPS compression algorithm, thus
the computational cost scaling becomes polynomially with $N_s$ which
allows much larger $N_s$. This new algorithm is called time-evolving
matrix product operators (TEMPO). The compression is achieved by
performing singular value decomposition on each tensor in MPS and
dispensing components with singular value below a threshold value. In
this article, the threshold value is $\varepsilon S_{\mathrm{max}}$,
where $\varepsilon$ is the control parameter and $S_{\mathrm{max}}$ is
the largest singular value. There is also another approach for tensor
network representation of discretized path integral
\cite{oshiyama2020-kibble,oshiyama2022-classical}.

We construct the tensor $K[s]F[s]$ as MPS using TEMPO method and
naturally represent tensors $X_{\xi_{\alpha}}[s]$ and
$Y_{\chi_{\alpha}}[s]$ as MPO, then formula
\eqref{eq:path-integral-rho}, \eqref{eq:generating-functional-xi} and
\eqref{eq:generating-functional-chi} can be easily evaluated.  We set
time interval $\delta t=0.02$ and the truncation parameter $N_s=200$
throughout this article. Here we choose a small value of $\delta t$,
which already exceeds the demand of convergence, in order to obtain
smooth curves in short time. The corresponding convergence analyses
are shown in \ref{app:convergence}.

\section{Spin-Boson Model with A Single Bath}
\label{sec:spin-boson-model-with-single-bath}
When evolution begins, system and bath start to correlate and
interaction energy arises. Let us first consider an unbiased
spin-boson model \cite{leggett1987-dynamics,weiss1993-quantum} with a
single bath to demonstrate the details of this process. The system
Hamiltonian is
\begin{equation}
  \hat{H}_d=\frac{\Delta}{2}\hat{\sigma}_x,
\end{equation}
where $\Delta$ is the tunneling amplitude between two states of the
spin.  Here we use $\Delta$ as the energy reference which fixes scales
of other parameters, and set its value as $\Delta=1$ for simplicity.
The bath Hamiltonian and the coupling term are
\begin{equation}
  \label{eq:single-bath-hamiltonian}
  \hat{H}_{\mathrm{E}}=\sum_k\omega_k\hat{b}_k^{\dag}\hat{b}_k,\quad \hat{H}_{d\mathrm{E}}=\hat{\sigma}_z\sum_kV_k(\hat{b}_k^{\dag}+\hat{b}_k).
\end{equation}
We set the spin to the $z$ direction at the initial time
$t=0$ for which
\begin{equation}
  \hat{\rho}_d(0)=\mqty(1&0\\0&0\\),\quad\expval{\hat{\sigma}_z(0)}=1,\quad\expval{\hat{\sigma}_x(0)}=0.
\end{equation}
There is only a single bath, and thus the heat that flows out from the
bath flows into the system and system-bath coupling part. The directly
evaluated quantities are the heat current out from the bath $I(t)$,
the system energy $E(t)$ and the interaction energy $W(t)$. The total
heat that flows out from the bath is simply calculated numerically as
$Q(t)=\sum_{i=0}^NI(t_i)\delta t$, and the change per unit time of the
system energy is also evaluated simply as
$\dot{E}(t_i)=[E(t_{i+1})-E(t_i)]/\delta t$. The quantity $\dot{W}(t)$
is evaluated in the same way.

\begin{figure}[htbp]
  \centerline{\includegraphics[]{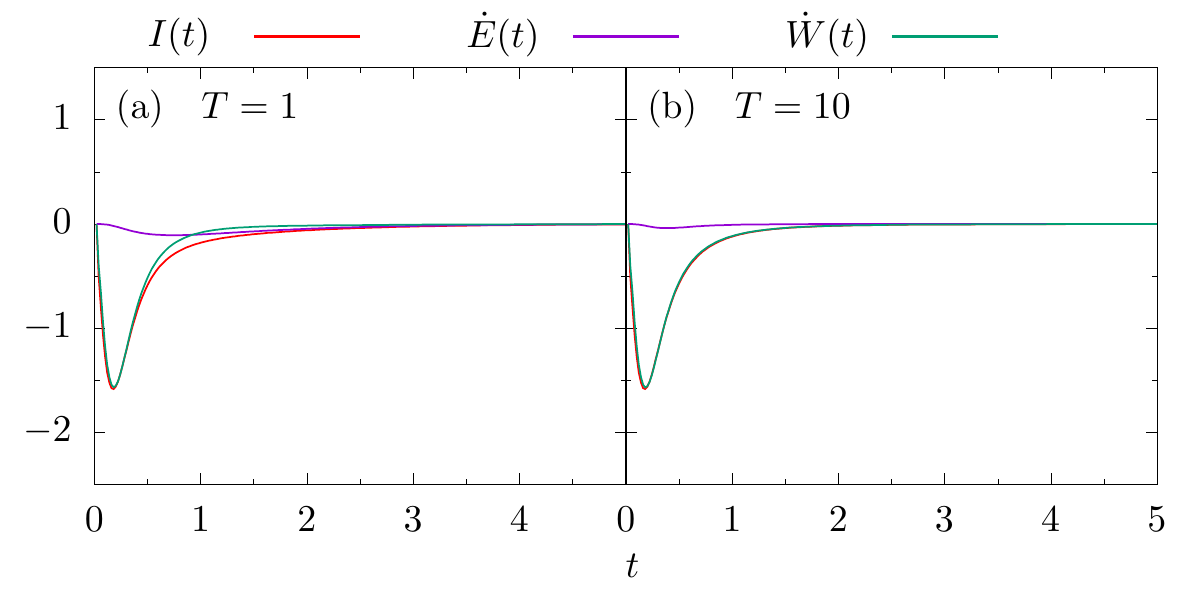}}
  \caption{The current $I(t)$ and time derivates $\dot{E}(t)$ and
    $\dot{W}(t)$ with coupling strength $\lambda=0.1$ at temperature
    (a) $T=1$ and (b) $T=10$.}
  \label{fig:02}
\end{figure}

\begin{figure}[htbp]
\centerline{\includegraphics[]{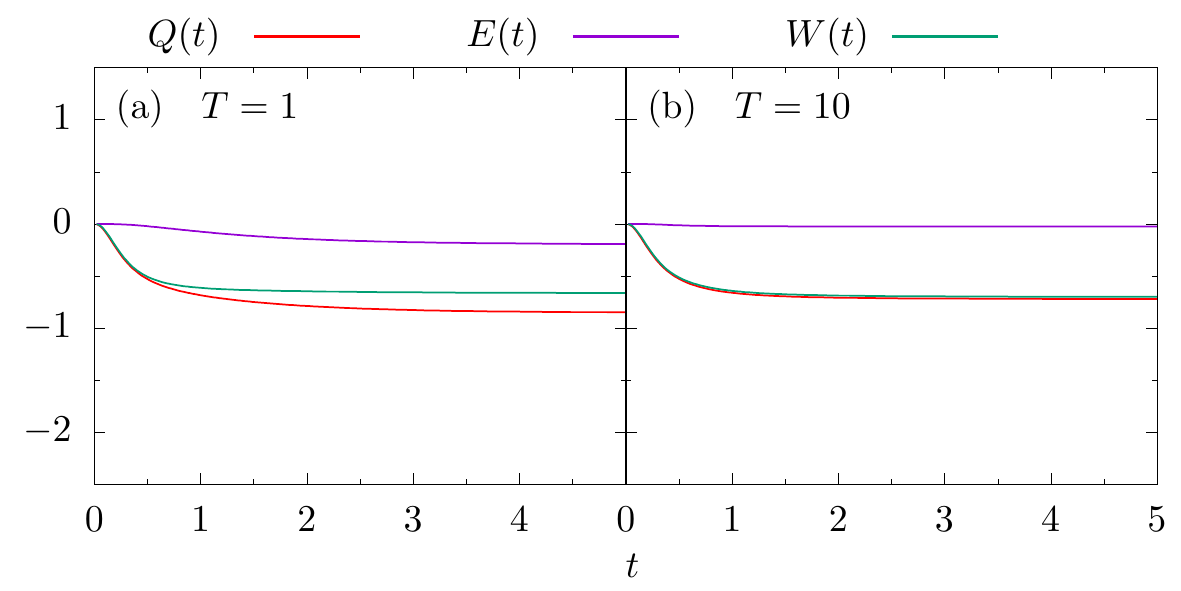}}
\caption{The heat that flows out from bath $Q(t)$, the system energy
  $E(t)$ and interaction energy $W(t)$ with coupling strength
  $\lambda=0.1$ at temperature (a) $T=1$ and (b) $T=10$.}
\label{fig:03}
\end{figure}

Set the system-bath coupling strength $\lambda=0.1$, the heat current
$I(t)$, along with time derivate quantities $\dot{E}(t),\dot{W}(t)$ at
different bath temperatures $T=1$ and $T=10$ are shown in
Fig. \ref{fig:02}. The heat that flows out from bath $Q(t)$, the
system energy $E(t)$ and the interaction energy $W(t)$ are shown in
Fig. \ref{fig:03}.

It can be seen from Fig. \ref{fig:02} that at the beginning of the
evolution, the current $I(t)$ and the quantity $\dot{W}(t)$ are both
negative and their values are close. They reach a minimum soon, then
tends to zero when approaching steady state. The magnitude of both
$I(t)$ and $\dot{W}(t)$ is much larger than $\dot{E}(t)$, this means
that the major energy change is not the heat transfer between system
and bath but the interaction energy change due to the correlation of
system and bath. In the asymptotic limit, transport quantities
$I(t),\dot{E}(t)$ and $\dot{W}(t)$ all tends to zero, as they should
be.

The system energy is just
$E(t)=\frac{\Delta}{2}\langle\hat{\sigma}_x(t)\rangle$, and it tends
to a negative value asymptotically at lower temperature $T=1$, as
shown in Fig. \ref{fig:03}(a). In this case, the energy that flows out
from the system becomes interaction energy and heat in the bath. Due
to this amount of energy transfer, $\dot{W}(t)$ is slightly larger
than $I(t)$, as shown in Fig. \ref{fig:02}(a).  At high temperature
$T=10$, the asymptotic value of $E(t)$ is close to zero
[Fig. \ref{fig:03}(b)], therefore there is not much energy flowing out
from the system. In this case, $\dot{W}(t)$ and $I(t)$ almost
coincide, as shown in Fig. \ref{fig:03}(b).

\section{Spin-Boson Model with Two Baths}
\label{sec:spin-boson-model-with-two-baths}

It is clear that there would be no steady current with a single bath,
thus it can be hardly called a transport problem. Now we consider a
spin-boson model with two baths, in which the steady current exists
when two baths are at different temperatures.  Let us label two baths
as 1st and 2nd. Correspondingly, the heat currents that flow out from
the baths are denoted as $I_1(t)$ and $I_2(t)$, and the total heats
that flow out from the baths are $Q_1(t)$ and $Q_2(t)$. The
corresponding interaction energies are denoted as $W_1(t)$ and
$W_2(t)$, and their time derivates are $\dot{W}_1(t)$ and $\dot{W}_2$.

\begin{figure}[htbp]
  \centerline{\includegraphics[]{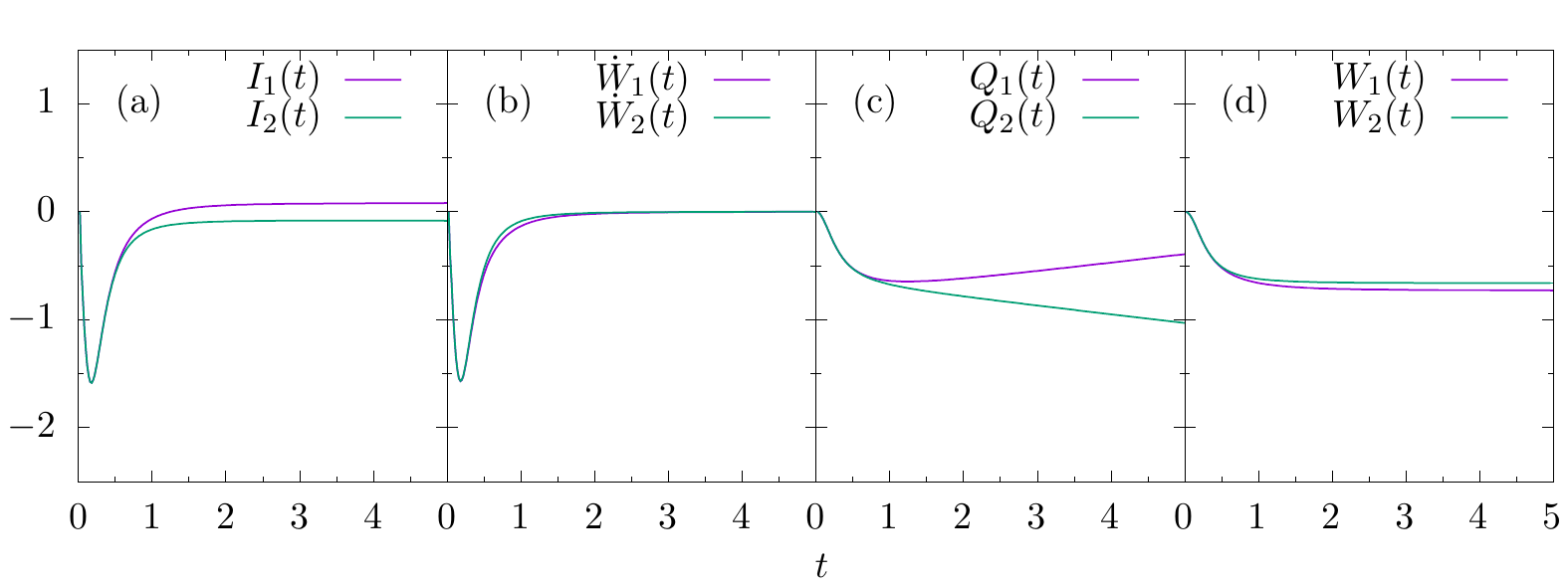}}
  \caption{Nonequilibrium transport process in two baths spin-boson
    model with same coupling strength $\lambda_1=\lambda_2=0.1$ but at
    different temperatures $T_1=10$ and $T_2=1$.}
\label{fig:04}
\end{figure}

We first consider the case where two baths are at different
temperatures for which the 1st bath is at high temperature $T_1=10$
and 2nd bath is at lower temperature $T_2=1$. The coupling strengths
between both baths are the same that $\lambda_1=\lambda_2=0.1$. The
results are shown in Fig. \ref{fig:04}.

We denote steady state quantities as their asymptotic limit
\begin{equation}
  \dot{E},\dot{W}_1,\dot{W}_2,I_1,I_2=
  \lim_{t\to\infty}\dot{E}(t),\dot{W}_1(t),\dot{W}_2(t),I_1(t),I_2(t).
\end{equation}
In steady state, the system energy remains constant and
correspondingly $\dot{E}=0$, therefore $E(t)$ and $\dot{E}(t)$ are not
shown in the figure. It can be seen that in steady state, the
interaction energies $W_1(t)$ and $W_2(t)$ are constant
[Fig. \ref{fig:04}(d)] and correspondingly $\dot{W}_1=\dot{W}_2=0$
[Fig. \ref{fig:04}(b)]. In this case, the heat that flows out from one
bath all flows into another bath, which indicates that a steady heat
current between the two baths is established.

Figure \ref{fig:04}(a) shows that in steady state $I_1$ is positive
while $I_2$ is negative. This means that heat flows out from 1st bath
to 2nd bath, i.e., flows from the bath of higher temperature to the
bath of lower temperature. We can also see in Fig. \ref{fig:04}(a)
that $I_1+I_2=0$, which expresses the energy conservation in steady
state.

Because of the existence of steady heat current, the total heat flows
out from baths $Q_{\alpha}(t)$ would not become constant, which can be
seen from Fig. \ref{fig:04}(c). In the beginning, both $Q_1(t)$ and
$Q_2(t)$ are negative, this is because that interaction energy and
system energy flow into baths. When steady state is established,
$Q_1(t)$ increases linearly with time and eventually becomes positive,
while $Q_2(t)$ decreases linearly.

\begin{figure}[htbp]
  \centerline{\includegraphics[]{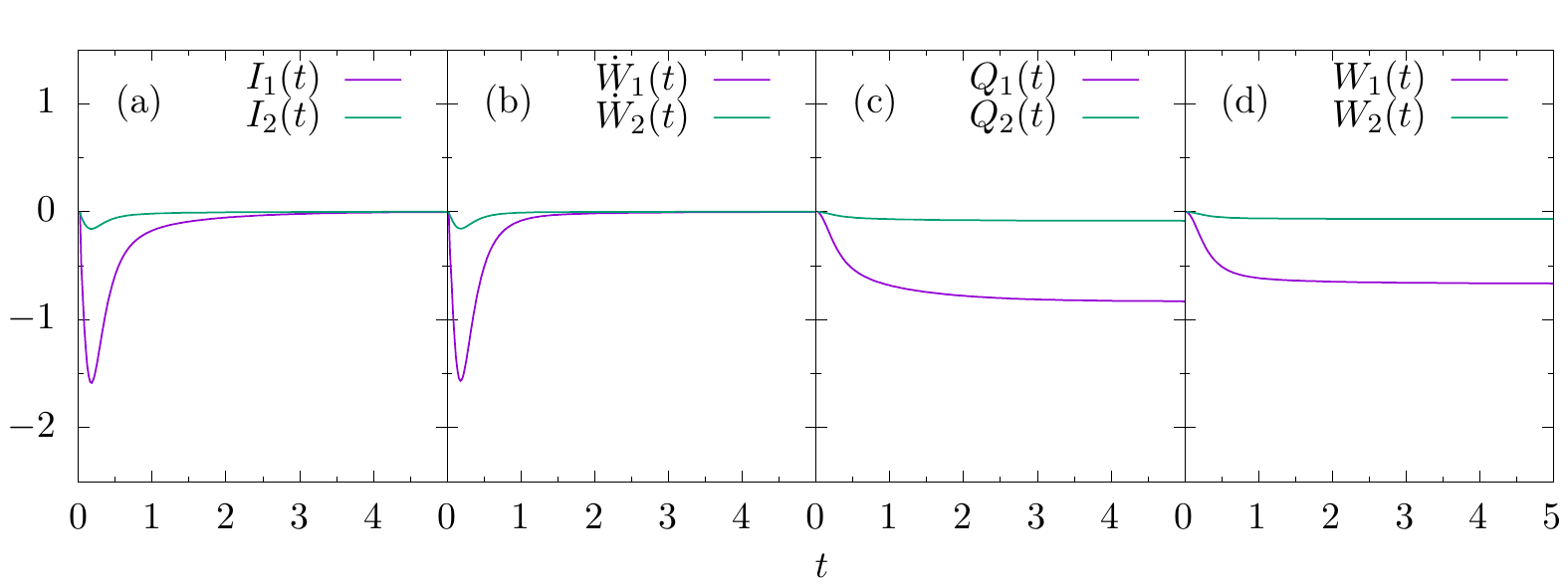}}
  \caption{Nonequilibrium transport process in spin-boson model with
    two baths at same temperature $T_1=T_2=1$ but with different
    coupling strengths $\lambda_1=0.1$ and $\lambda_2=0.01$.}
\label{fig:05}
\end{figure}

Figure \ref{fig:05} shows the results when two baths are at the same
temperature but with different coupling strengths. The parameters are
$\lambda_1=0.1$, $\lambda_2=0.01$ and $T_1=T_2=1$. In this case,
$\lambda_1$ is much larger than $\lambda_2$, and thus in the beginning
the magnitude of $I_1(t)$ and $\dot{W}_1(t)$ are much larger than that
of $I_2(t)$ and $\dot{W}_2(t)$ since much more energy can be extracted
from $W_1(t)$. When arriving at steady state, both $I_1(t)$ and
$I_2(t)$ tends to zero which shows that no steady current exists
between two baths at the same temperature.

In a quantum transport problem with two baths, the observable of
interesting is often the symmetrized current
\begin{equation}
  I(t)=\frac{1}{2}[I_1(t)-I_2(t)].
\end{equation}
The steady state current is denoted as $I=\lim_{t\to\infty}I(t)$.

\begin{figure}[htbp]
  \centerline{\includegraphics[]{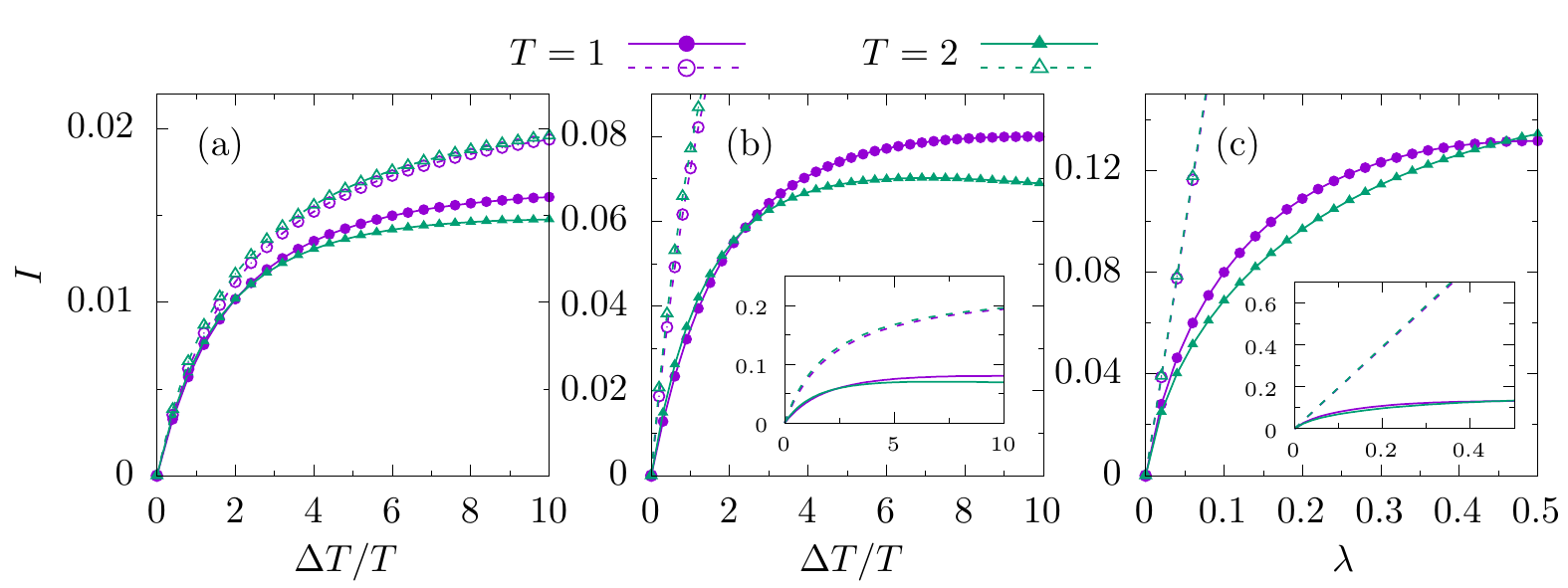}}
  \caption{Steady symmetrized currents with base temperatures $T=1$
    and $T=2$, the solid lines correspond to TEMPO results and the
    dashed lines correspond to BMME results. (a) Currents with respect
    to $\Delta T/T$ with weak coupling $\lambda=0.01$. (b) Currents
    with respect to $\Delta T/T$ with larger coupling $\lambda=0.1$,
    and the inset shows the same data in a larger scale. (c) Currents
    with respect to coupling strength $\lambda$ with $\Delta T/T=10$,
    and the inset shows the same data in a larger scale.}
\label{fig:06}
\end{figure}

The steady current is determined by the temperature bias between two
baths. For convenience, we denote $T=T_2$ as the base temperature and
$\Delta T=T_1-T_2$ as temperature bias. We set the coupling strengths
of both baths to be equal for which $\lambda_1=\lambda_2=\lambda$.
The steady currents beyond the linear response regime with different
base temperatures $T=1$ and $T=2$ are shown in Fig. \ref{fig:06}. Here
we compare our results to that by standard Born-Markov master equation
(BMME) \cite{breuer2007-the}, whose details refer to
\ref{app:master-equation}. In the figure, the TEMPO results are shown
in solid lines and the BMME results are shown in dashed lines.
Fig. \ref{fig:06}(a) shows currents with respect to $\Delta T/T$ with
weak coupling ($\lambda=0.01$), and Fig. \ref{fig:06}(b) shows results
with larger coupling $\lambda=0.1$. In Fig. \ref{fig:06}(c), the
currents with respect to coupling strength $\lambda$ with
$\Delta T/T=10$ are shown.

Let us first look at Figs. \ref{fig:06}(a) and \ref{fig:06}(b). In a
slightly nonequilibrium situation, i.e., linear response regime, the
steady heat current is supposed to be proportional to $\Delta T/T$. It
can be seen that it is indeed the case when $\Delta T/T$ is small,
while when $\Delta T/T$ becomes larger the linear response
approximation fails. The magnitude of $I$ will saturate when
$\Delta T/T$ is large enough.  For both weak and larger coupling
cases, BMME results with $T=1$ and $T=2$ almost coincide. This means
that within the Markovian approximation, the steady current almost
depends on only $\Delta T/T$ and the absolute value of $T$ is not
important. On the contrary, TEMPO results, which fully take the
non-Markovian effects, show that the steady currents with different
base temperatures $T$ would deviate as $\Delta T/T$ increases.

Within the Markovian approximation, the dependence of current on
temperature bias is always monotonic for which larger $\Delta T/T$
would induce larger current $I$. However, the $T=2$ TEMPO result shows
that with larger coupling strength $\lambda=0.1$
[Fig. \ref{fig:06}(b)], the steady current shows nonmonotonic
dependence of temperature bias for which $I$ can even slightly
decrease with increasing $\Delta T/T$. This is clearly a consequence
of the non-Markovian effect.

In weak coupling case $\lambda=0.01$ [Fig. \ref{fig:06}(a)], the
difference between TEMPO results and BMME results is small even with
large $\Delta T/T$. This is not surprising since BMME should work in
the weak coupling limit. When coupling strength is increased by 10
times that $\lambda=0.1$ [Fig. \ref{fig:06}(b)], the BMME currents
also increase nearly 10 times. It can be seen from
Fig. \ref{fig:06}(c) that within Markovian approximation, the current
is linear to system-bath coupling, therefore increasing coupling
strength can significantly enhance the current. However, TEMPO results
show that when considering non-Markovian effects, the enhancement by
increasing coupling is not that significant for which the magnitude of
current would saturate when $\lambda$ is large enough. This indicates
that when designing nano devices the improvement of thermal
conductivity by simply increasing the coupling strength is limited.
\section{Conclusions}
\label{sec:conclusions}

Iterative tensor multiplication based on path integral formalism is an
important method to simulate reduced dynamics of quantum open
systems. For fermionic environments, the corresponding tensor can be
numerically obtained via the determinant formula
\cite{segal2010-numerically,agarwalla2017-anderson,chen2019-dissipative,chen2020-landau}.
For system linearly coupled to bosonic environments, the influence
functional can be analytically represented by Feynman-Vernon formalism
\cite{feynman1963-the,weiss1993-quantum,caldeira1983-path}, and after
discretization we obtain the corresponding tensor. The discretization
scheme employed is usually the QUAPI method
\cite{makarov1993-tunneling,makri1995-numerical,dattani2012-analytic}.

To handle quantum transport problems in path integral formalism, the
knowledge of generating functional is necessary. For fermionic
environment, the generating functional can be obtained via the
determinant formula and the iterative tensor multiplication scheme,
then the current is obtained by a numerical differentiation
\cite{weiss2008-iterative,segal2011-nonequilibrium,bedkihal2012-dynamics,simine2013-path}.
The iterative tensor multiplication approach is numerical exact and
fully non-Markovian, but its computational cost scales
exponentially. Such a problem can be resolved for the system linearly
coupled to the bosonic environment by TEMPO algorithm, where the
tensor obtained via QUAPI is represented by MPS
\cite{strathearn2018-efficient,gribben2022-exact}.

In this article, we generalize TEMPO algorithm to nonequilibrium
quantum transport problems. The generating functional is evaluated as
a tensor network via QUAPI and TEMPO method, and its numerical
differentiation gives the corresponding heat current. A key point in
this evaluation scheme is that the auxiliary variable in generating
functional needs to be handled in the same way as the system
variable. That is, it should be treated as a path segment rather than
a single variable and needs to be also discretized via QUAPI scheme.

A part of the heat that flows out from one bath flows into the system
and other baths, and the rest becomes the interaction energy, i.e.,
the energy stored in the system-bath coupling part. The spin-boson
model is used to demonstrate the details of such heat flowing. In
addition, steady currents between two baths at different temperatures
beyond the linear response regime are demonstrated.

\section*{Acknowledgement.}
This work is supported by the NSFC Grant No. 12104328.

\appendix
\section{Derivation of Path Integral Formalism}
\label{app:details-of-path-integral-formalsim}
Here we give a detailed derivation of the path integral formalism.
For simplicity, let us first consider the situation with a single bath
consisting of a single boson mode. The generalization to continuous
boson modes and multiple baths is straightforward. In this case, the
bath Hamiltonian and the system-bath coupling are
\begin{equation}
  \hat{H}_{\mathrm{E}}=\omega \hat{b}^{\dag}\hat{b},\quad \hat{H}_{d\mathrm{E}}=\hat{\sigma}_z V(\hat{b}^{\dag}+\hat{b}).
\end{equation}
The bath can be described by boson coherent
states \cite{negele1998-quantum,zhang1990-coherent} which is defined
as the eigenstate of $b$ that
\begin{equation}
  \hat{b}\ket{\varphi}=\varphi\ket{\varphi},\quad
  \bra{\varphi}\hat{b}^{\dag}=\bra{\varphi}\bar{\varphi},
\end{equation}
where $\varphi$ is a complex number and $\bar{\varphi}$ is its complex
conjugate. The identity operator in bath Hilbert space can be
expressed in terms of coherent states as
\begin{equation}
  \int\frac{\dd{\bar{\varphi}}\dd{\varphi}}{2\pi i}e^{-\bar{\varphi}\varphi}\ketbra{\varphi}{\varphi}.
\end{equation}
Accordingly, the trace over the bath can be written as
\begin{equation}
  \Tr_{\mathrm{E}}[\cdots]=\int\frac{\dd{\bar{\varphi}}\dd{\varphi}}{2\pi i} e^{-\bar{\varphi}\varphi}
  \mel{\varphi}{\cdots}{\varphi}.
\end{equation}
Suppose states $\ket{s}$ spans the Hilbert space of the system, i.e.,
$\sum_s\ketbra{s}{s}$ is the identity operator in system Hilbert
space, then the trace over the system can be expressed as
\begin{equation}
  \Tr_d[\cdots]=\sum_s\mel{s}{\cdots}{s}.
\end{equation}
Correspondingly, the identity operator in the whole Hilbert space can
be written as
\begin{equation}
  \label{eq:identity}
  1=\sum_s\int \frac{\dd{\bar{\varphi}}\dd{\varphi}}{2\pi i}e^{-\bar{\varphi}\varphi}
  \ketbra{s\varphi}{s\varphi}.
\end{equation}
Insert the above identity operator every time step in
\eqref{eq:rho-split} and employ the first order Trotter-Suzuki
decomposition \cite{trotter1959-product,suzuki1976-generalized} that
\begin{equation}
  e^{\mp i\delta t\hat{H}}=e^{\mp i\delta t\hat{\sigma}_zV\hat{b}^{\dag}}e^{\mp i\delta t\hat{H}_d}
  e^{\mp i\delta t\hat{H}_{\mathrm{E}}}e^{\mp i\delta t\hat{\sigma}_zV\hat{b}},
\end{equation}
and we have
\begin{equation}
    \mel{s_{\alpha+1}^+\varphi_{\alpha+1}^+}{e^{-i\delta t\hat{H}}}{s_{\alpha}^+\varphi_{\alpha}^+}
    =\mel{s_{\alpha+1}^+}{e^{-i\delta t\hat{H}_d}}{s_{\alpha}^+}
    e^{a\bar{\varphi}_{\alpha+1}^+\varphi_{\alpha}^+-i\delta tV(s_{\alpha+1}^+\bar{\varphi}_{\alpha+1}^++s_{\alpha}^+\varphi_{\alpha}^+)},
\end{equation}
and
\begin{equation}
    \mel{s_{\alpha}^-\varphi_{\alpha}^-}{e^{i\delta t\hat{H}}}{s_{\alpha+1}^-\varphi_{\alpha+1}^-}
    =\mel{s_{\alpha}^-}{e^{i\delta t\hat{H}_d}}{s_{\alpha+1}^-}
    e^{a\bar{\varphi}_{\alpha}^-\varphi_{\alpha+1}^-+i\delta tV(s_{\alpha}^-\bar{\varphi}_{\alpha}^-+s_{\alpha+1}^-\varphi_{\alpha+1}^-)}.
\end{equation}
The element of the reduced density matrix is then
\begin{equation}
  \rho_d(s_N^{\pm})=\sum_{s_0^{\pm},\ldots,s_{N-1}^{\pm}}K(s_0^{\pm},\ldots,s_N^{\pm})F(s_0^{\pm},\ldots,s_N^{\pm}),
\end{equation}
where $K(s_0^{\pm},\ldots,s_N^{\pm})$ is the bare system propagator
\begin{equation}
  K(s_0^{\pm},\ldots,s_N^{\pm})=\mel{s_N^+}{e^{-iH_d\delta t}}{s_{N-1}^+}\cdots\mel{s_0^+}{\hat{\rho}_d(0)}{s_0^-}\cdots\mel{s_{N-1}^-}{e^{iH_d\delta t}}{s_N^-}.
\end{equation}

Define two vectors of $2N-1$ elements as
\begin{equation}
  \bar{\varsigma}=(s_0^+,\ldots,s_{N-1}^+,s_N^-,s_{N-1}^-,\ldots,s_0^-),
\end{equation}
and
\begin{equation}
  \varsigma=(s_0^+,\ldots,s_{N-1}^+,s_N^+,s_{N-1}^-,\ldots,s_0^-).
\end{equation}
Note that here $\bar{\varsigma}$ is not the Hermitian conjugate of
$\varsigma$, the bar over $\bar{\varsigma}$ is just a symbol for
convenience. Besides, we define a time interval operator on the
contour $\dd{t}$ as $\delta t$ when acting on $s$ on the forward
branch and $-\delta t$ on the backward branch. Due to the trace over
the bath, we have $\varphi(t_N^+)=\bar{\varphi}_N$ and
$\varphi(t_N^-)=\varphi_N$. Denote a vector
$\varphi=(\varphi_0^+,\ldots,\varphi_{N-1}^+,\varphi_N,\varphi_{N-1}^-,\ldots,\varphi_0^-)$
and $\bar{\varphi}$ as its Hermitian conjugate, the influence
functional $F(s_0^{\pm},\ldots,s_N^{\pm})$ can be written as
\begin{equation}
  \int D[\bar{\varphi}\varphi]e^{-\sum_{j=0}^{2N-1}\sum_{k=0}^{2N-1}\bar{\varphi}_jS_{jk}\varphi_k-
    i\dd{t}V\sum_{k=0}^{2N-1}(\bar{\varsigma}_k\varphi_k+\varsigma_k\bar{\varphi}_k)},
\end{equation}
where 
\begin{equation}
  D[\bar{\varphi}\varphi]=
  \prod_{k=1}^{2N-1}\frac{\dd{\bar{\varphi}_k\dd{\varphi_k}}}{2\pi i}.
\end{equation}
This expression can be written in a more compact form as
\begin{equation}
  \label{eq:influence-functional-before-gaussian}
  \int D[\bar{\varphi}\varphi]e^{-\bar{\varphi}S\varphi-i\dd{t}V(\bar{\varsigma}\varphi+\varsigma\bar{\varphi})},
\end{equation}
where $S$ is a $(2N-1)\times(2N-1)$ matrix
\begin{equation}
  S=\mqty(
  1 & & & & & & e^{-\omega/T}\\
  -a & 1 & & & & & \\
  & -a & \ddots & & & & \\
  & & & 1 & & & \\
  & & & -\bar{a}&\ddots & & \\
  & & & & & 1 & \\
  & & & & & -\bar{a} & 1\\
  ),
\end{equation}
where $a=e^{-i\delta t\omega}$ and $\bar{a}=e^{i\delta t\omega}$. The
expression \eqref{eq:influence-functional-before-gaussian} can carried
out via Gaussian integral as
\begin{equation}
  \label{eq:influence-functional-after-gaussian}
  F(s_0^{\pm},\ldots,s_N^{\pm})=[\det S]^{-1}e^{-V^2\dd{t}\bar{\varsigma}S^{-1}\dd{t}\varsigma}.
\end{equation}
The determinant of $S$ is just $1-e^{-\omega/T}$ and then
$[\det S]^{-1}$ gives the partition function of the free bath. The
element of inverse matrix $S_{ij}^{-1}$ is just the contour Green's
function for which
\begin{equation}
  S^{-1}_{ij}=G(t_i,t_j),
\end{equation}
where
$G(t_i,t_j)=\langle
T_{\mathcal{C}}\hat{b}(t_i)\hat{b}^{\dag}(t_j)\rangle_0$. Therefore in the
continuous limit, the influence functional can be written as
\begin{equation}
  F[s]=e^{-\int_{\mathcal{C}}\dd{t'}\int_{\mathcal{C}}\dd{t''}s(t')\varDelta(t',t'')s(t'')},
\end{equation}
where
\begin{equation}
  \varDelta(t',t'')=V^2G(t',t'').
\end{equation}

Now let us evaluate the generating functional
\begin{equation}
  \hat{\rho}_d^{\xi}(t)=\Tr_{\mathrm{E}}[\hat{\rho}(t)e^{\xi V\omega \hat{b}}].
\end{equation}
With such an extra source term, the influence functional
\eqref{eq:influence-functional-before-gaussian} before Gaussian
integral becomes
\begin{equation}
  \int D[\bar{\varphi}\varphi]
  e^{-\bar{\varphi}S\varphi-i\dd{t}V(\bar{\varsigma}\varphi+\varsigma\bar{\varphi})+\xi V\omega\varphi_N},
\end{equation}
then after the Gaussian integral we have an extra term
\begin{equation}
    X_{\xi}[s]=e^{-i\xi\int_{\mathcal{C}}\dd{t'}s(t')\varGamma(t',t_N^-)},
\end{equation}
where $\varGamma(t',t'')=\omega V^2G(t',t'')$. For generating
functional
$\hat{\rho}^{\chi}_d(t)=\Tr_{\mathrm{E}}[\hat{\rho}(t)e^{\chi  V\hat{b}}]$,
we can apply the same procedure and obtain an extra term
\begin{equation}
  Y_{\chi}[s]=e^{-i\chi\int_{\mathcal{C}}\dd{t'}s(t')\varDelta(t',t_N^-)}.
\end{equation}

Now let us turn to the situation with a single bath consisting of
continuous boson modes. In this case, the bath Hamiltonian and the
system-bath coupling are same to
\eqref{eq:single-bath-hamiltonian}. Following the same procedure, we
shall find that the influence functional now writes
$F[s]=\prod_kF_k[s]$, where
\begin{equation}
  F_k[s]=e^{-\int_{\mathcal{C}}\dd{t'}\int_{\mathcal{C}}\dd{t''}s(t')\varDelta_k(t',t'')s(t'')}
\end{equation}
with
[$G_k(t',t'')=\langle T_{\mathcal{C}}\hat{b}_k(t')\hat{b}_k^{\dag}(t'')\rangle_0$]
\begin{equation}
  \varDelta_k(t',t'')=V_k^2G_k(t',t'').
\end{equation}
Therefore
\begin{equation}
    F[s]=e^{-\sum_k\int_{C}\dd{t'}\int_{\mathcal{C}}\dd{t''}s(t')\varDelta_k(t',t'')s(t'')}
    =e^{-\int_{C}\dd{t'}\int_{\mathcal{C}}\dd{t''}s(t')\varDelta(t',t'')s(t'')},
\end{equation}
where
\begin{equation}
  \varDelta(t',t'')=\sum_kV_k^2G_k(t',t'')=\int\dd{\omega}J(\omega)G_{\omega}(t',t'').
\end{equation}
Here $G_{\omega}(t',t'')$ is $G_k(t',t'')$ when
$\omega_k=\omega$. Similarly, we have
\begin{equation}
  X_{\xi}[s]=e^{-i\xi\int_{\mathcal{C}}\dd{t'}s(t')\varGamma(t',t_N^-)},
\end{equation}
where
\begin{equation}
  \varGamma(t',t'')=\sum_k\omega_kV_k^2G_k(t',t'')=\int\dd{\omega}\omega J(\omega)G_{\omega}(t',t''),
\end{equation}
and
\begin{equation}
  Y_{\chi}[s]=e^{-i\chi\int_{\mathcal{C}}\dd{t'}s(t')\varDelta(t',t_N^-)}.
\end{equation}

The extension to multiple baths is the same as what we have done
above, and finally we shall obtain the path integral formulas shown in
Sec. \ref{sec:path-integral-formalism}.

\section{Discretization Scheme}
\label{app:discretization-scheme}
Here we represent the details of the discretization scheme of the path
integral formalism. The specific form of the contour ordered Green's
function are
\begin{equation}
  G_{\alpha k}^{++}(t'-t'')=\begin{cases}
    (1+n_{\alpha k})e^{-i\omega_{\alpha k}(t'-t'')}, & t'\ge t'';\\
    n_{\alpha k}e^{-i\omega_{\alpha k}(t'-t'')}, & t'< t'',\\
  \end{cases}
\end{equation}
\begin{equation}
  G_{\alpha k}^{--}(t'-t'')=\begin{cases}
    n_{\alpha k}e^{-i\omega_{\alpha k}(t'-t'')}, & t'\ge t'';\\
    (1+n_{\alpha k})e^{-i\omega_{\alpha k}(t'-t'')} & t'<t'',
  \end{cases}
\end{equation}
\begin{equation}
  G_{\alpha k}^{+-}(t'-t'')=n_{\alpha k}e^{-i\omega(t'-t'')},
\end{equation}
\begin{equation}
  G_{\alpha k}^{-+}(t'-t'')=(1+n_{\alpha k})e^{-i\omega_{\alpha k}(t'-t'')},
\end{equation}
where $n_{\alpha k}=(e^{\omega_{\alpha k}/T_{\alpha}}-1)^{-1}$ is the
Bose-Einstein distribution function. Substituting the above specific
form of $G$ into \eqref{eq:F} and noticing that the area integral
$\int_0^t\dd{t'}\int_0^t\dd{t''}$ can be split as
$\int_{t'\ge t''}\dd{t'}\dd{t''}$ and $\int_{t'<t''}\dd{t'}\dd{t''}$,
we shall obtain a specific form of $F_{\alpha}[s]$ as
\eqref{eq:influence-functional}. In order to discretize the influence
functional $F_{\alpha}[s]$, we need to discretize the autocorrelation
function \eqref{eq:autocorrelation} according to formula
\eqref{eq:eta-jk} and \eqref{eq:eta-jj}.

Now let us turn to the generating functional term
$X_{\xi_{\alpha}}[s]$, whose path integral formalism is
\eqref{eq:X}. The second argument in $\varGamma_{\alpha}(t',t_N^-)$ is
fixed at the starting point of the backward branch, therefore when
$t'$ is on the forward branch we have $\varGamma_{\alpha}^{+-}$ and
when $t'$ is on the backward branch we have
$\varGamma_{\alpha}^{--}$. The contour integral in \eqref{eq:X} is
thus
\begin{equation}
  \int_{\mathcal{C}}\dd{t'}s_{\alpha}(t')\varGamma_{\alpha}(t',t_N^-)
  =\int_0^t\dd{t'}[s_{\alpha}^+(t')\varGamma_{\alpha}^{+-}(t',t)-s_{\alpha}^-(t')\varGamma_{\alpha}^{--}(t',t)].
\end{equation}
Applying QUAPI scheme to above formula, we shall obtain the
discretization formula \eqref{eq:X-discretization}.

Similar to the situation of $X_{\xi_{\alpha}}[s]$, the contour
integral in $Y_{\xi_{\alpha}}[s]$, whose path integral formalism is
\eqref{eq:Y}, is
\begin{equation}
  \int_{\mathcal{C}}\dd{t'}s_{\alpha}(t')\varDelta_{\alpha}(t',t_N^-)
  =\int_0^t\dd{t'}[s_{\alpha}^+(t')\varDelta_{\alpha}^{+-}(t',t)-s^-(t')\varDelta_{\alpha}^{--}(t',t)].
\end{equation}
Therefore after the discretization, we have
\begin{equation}
  Y_{\chi_{\alpha}}[s]=e^{-i\chi_{\alpha}\sum_{j=0}^Ns^+_{\alpha j}\zeta^{+-}_{jN}-s^-_{\alpha j}\zeta_{jN}^{--}},
\end{equation}
where for $j\ne N$
\begin{equation}
  \zeta_{jN}^{+-}=\frac{1}{\delta t}\int_{(N-\frac{1}{2})\delta t}^{(N+\frac{1}{2})\delta t}\dd{t'}\
  \int_{(j-\frac{1}{2})\delta t}^{(j+\frac{1}{2})\delta t}\dd{t''}\varDelta_{\alpha}^{+-}(t''-t'),
\end{equation}
\begin{equation}
  \zeta_{jN}^{--}=\frac{1}{\delta t}\int_{(N-\frac{1}{2})\delta t}^{(N+\frac{1}{2})\delta t}\dd{t'}\
  \int_{(j-\frac{1}{2})\delta t}^{(j+\frac{1}{2})\delta t}\dd{t''}\varDelta_{\alpha}^{--}(t''-t'),
\end{equation}
and for $j=N$
\begin{equation}
  \zeta_{NN}^{+-}=\frac{1}{\delta t}\int_{(N-\frac{1}{2})\delta t}^{(N+\frac{1}{2})\delta t}\dd{t'}
  \int^{t'}_{(N-\frac{1}{2})\delta t}\dd{t''}\varDelta_{\alpha}^{+-}(t''-t'),
\end{equation}
\begin{equation}
  \zeta_{NN}^{--}=\frac{1}{\delta t}\int_{(N-\frac{1}{2})\delta t}^{(N+\frac{1}{2})\delta t}\dd{t'}
  \int^{t'}_{(N-\frac{1}{2})\delta t}\dd{t''}\varDelta_{\alpha}^{--}(t''-t').
\end{equation}

Note that here $t'$ is always not greater than $t$, according to
\eqref{eq:varGamma-G} the specific form of $\varGamma^{+-}_{\alpha}$
and $\varGamma_{\alpha}^{--}$ are 
\begin{equation}
  \varGamma_{\alpha}^{+-}(t'-t)=\int\dd{\omega}\omega J(\omega)n_{\alpha}(\omega)e^{-i\omega (t'-t)},
\end{equation}
and
\begin{equation}
  \varGamma_{\alpha}^{--}(t'-t)=\int\dd{\omega}\omega J(\omega)[1+n_{\alpha}(\omega)]e^{-i\omega (t'-t)},
\end{equation}
where $n_{\alpha}(\omega)=(e^{\omega/T_{\alpha}}-1)^{-1}$ is the
Bose-Einstein distribution function for $\alpha$th bath. Similarly,
according to \eqref{eq:varDelta-G}, the specific form of
$\varDelta_{\alpha}^{+-}$ and $\varDelta_{\alpha}^{--}$ are
\begin{equation}
  \varDelta_{\alpha}^{+-}(t'-t)=\int\dd{\omega}J(\omega)n_{\alpha}(\omega)e^{-i\omega (t'-t)},
\end{equation}
and
\begin{equation}
  \varDelta_{\alpha}^{--}(t'-t)=\int\dd{\omega}J(\omega)[1+n_{\alpha}(\omega)]e^{-i\omega (t'-t)}.
\end{equation}
At last, we list explicit formulas for these coefficients after the
discretization as
\begin{equation}
  \eta_{jk}=2\int_0^{\infty}\dd{\omega}\frac{J_{\alpha}(\omega)}{\omega^2}
  \qty[\coth\frac{\omega}{2T_{\alpha}}\cos[\omega(j-k)\delta t]-
  i\sin\omega(j-k)\delta t](1-\cos\omega\delta t),\quad j\ne k;
\end{equation}
\begin{equation}
  \eta_{jj}=\int_0^{\infty}\dd{\omega}\frac{J_{\alpha}(\omega)}{\omega^2}
  \qty[\coth\frac{\omega}{2T_{\alpha}}(1-\cos\omega\delta t)-
  i(\omega\delta t-\sin\omega\delta t)],
\end{equation}
\begin{equation}
  \gamma^{+-}_{jN}=2\int_0^{\infty}\dd{\omega}\frac{J_{\alpha}(\omega)}{\omega}n_{\alpha}(\omega)e^{-i\omega(j-N)\delta t}(1-\cos\omega\delta t),
  \quad j\ne N,
\end{equation}  
\begin{equation}
  \gamma^{--}_{jN}=2\int_0^{\infty}\dd{\omega}\frac{J_{\alpha}(\omega)}{\omega}[1+n_{\alpha}(\omega)]e^{-i\omega(j-N)\delta t}(1-\cos\omega\delta t),
  \quad j\ne N,
\end{equation}
\begin{equation}
  \gamma^{+-}_{NN}=\int_0^{\infty}\dd{\omega}\frac{J_{\alpha}(\omega)}{\omega}n_{\alpha}(\omega)[(1+i\omega\delta t)-e^{i\omega\delta t}],
\end{equation}
\begin{equation}
  \gamma^{--}_{NN}=\int_0^{\infty}\dd{\omega}\frac{J_{\alpha}(\omega)}{\omega}[1+n_{\alpha}(\omega)][(1+i\omega\delta t)-e^{i\omega\delta t}],
\end{equation}
\begin{equation}
  \zeta_{jN}^{+-}=2\int_0^{\infty}\dd{\omega}\frac{J_{\alpha}(\omega)}{\omega^2}n_{\alpha}(\omega)e^{-i\omega(j-N)\delta t}(1-\cos\omega\delta t),\quad j\ne N,
\end{equation}  
\begin{equation}
  \zeta_{jN}^{--}=2\int_0^{\infty}\dd{\omega}\frac{J_{\alpha}(\omega)}{\omega^2}[1+n_{\alpha}(\omega)]e^{-i\omega(j-N)\delta t}(1-\cos\omega\delta t),\quad j\ne N,
\end{equation}
\begin{equation}
  \zeta^{+-}_{NN}=\int_0^{\infty}\dd{\omega}\frac{J_{\alpha}(\omega)}{\omega^2}n_{\alpha}(\omega)[(1+i\omega\delta t)-e^{i\omega\delta t}],
\end{equation}
\begin{equation}
  \zeta^{--}_{NN}=\int_0^{\infty}\dd{\omega}\frac{J_{\alpha}(\omega)}{\omega^2}[1+n_{\alpha}(\omega)][(1+i\omega\delta t)-e^{i\omega\delta t}].
\end{equation}

\section{Born-Markov Master Equation}
\label{app:master-equation}
In Sec. \ref{sec:spin-boson-model-with-two-baths} we use Born-Markov
master equation to benchmark our steady current results. For the
spin-boson model with two baths in this article, the Born-Markov
master equation can be written as \cite{breuer2007-the}
\begin{equation}
  \pdv{t}\hat{\rho}_d(t)=-i[\hat{H}_d,\hat{\rho}_d(t)]+\mathcal{D}_1[\hat{\rho}(t)]+\mathcal{D}_2[\hat{\rho}(t)],
\end{equation}
where $\mathcal{D}_\alpha$ is the dissipator of $\alpha$th bath for which
\begin{equation}
  \mathcal{D}_{\alpha}[\hat{\rho}_d(t)]=-\int_0^{\infty}[\hat{\sigma}_z,\hat{\sigma}_z(-\tau)\hat{\rho}_d(t)]C_{\alpha}(\tau)\dd{\tau}+
  \int_0^{\infty}[\hat{\sigma}_z,\hat{\rho}_d(t)\hat{\sigma}_z(-\tau)]C_{\alpha}(-\tau)\dd{\tau}.
\end{equation}
Here $\hat{\sigma}_z(t)$ is the interaction picture operator
\begin{equation}
  \hat{\sigma}_z(t)=e^{i\hat{H}_dt}\hat{\sigma}_ze^{-i\hat{H}_dt}=
  \mqty(\cos\Delta t&-i\sin\Delta t\\
  i\sin\Delta t & -\cos\Delta t),
\end{equation}
and $C_{\alpha}(t)$ is the autocorrelation function of $\alpha$th bath
\begin{equation}
  C_{\alpha}(t)=\int J(\omega)\{n_{\alpha}(\omega)e^{i\omega t}+[1+n_{\alpha}(\omega)]e^{-i\omega t}\}\dd{\omega}.
\end{equation}

The energy change of the system is
\begin{equation}
  \begin{split}
    \dot{E}(t)=&\Tr_d[\hat{H}_d\pdv{\hat{\rho}_d(t)}{t}]\\
    =&-i\Tr_d\{\hat{H}_d[\hat{H}_d,\hat{\rho}_d(t)]\}+
       \Tr_d\{\hat{H}_d\mathcal{D}_1[\hat{\rho}_d(t)]\}+\Tr_d\{\hat{H}_d\mathcal{D}_2[\hat{\rho}_d(t)]\}.\\
  \end{split}
\end{equation}
Here the first term is zero, and the last two terms correspond to the
energy change due to baths, i.e., the currents flow into the system
from the baths. Therefore we have
$I_{\alpha}(t)=\Tr_d\{\hat{H}_d\mathcal{D}_{\alpha}[\hat{\rho}_d(t)]\}$.

Within Born-Markov approximation, there is no interaction energy
$W(t)$ since the density matrix is always decoupled into system and
bath parts, and thus we only consider the steady state currents. In
steady state, the system energy change $\dot{E}=0$ and then we have
$I_1+I_2=0$.

For evaluation of master equation, we define a quantity
\begin{equation}
  \begin{split}
    C_{\alpha}(E)=&\int_0^{\infty}e^{iEt}C_{\alpha}(t)\dd{t}\\
    =&\pi\{J_{\alpha}(-E)n_{\alpha}(-E)+J_{\alpha}(E)[1+n_{\alpha}(E)]\}+
    iP\int J_{\alpha}(\omega)\qty[\frac{n_{\alpha}(\omega)}{E+\omega}+\frac{1+n_{\alpha}(\omega)}{E-\omega}]\dd{\omega},
\end{split}
\end{equation}
where $P$ denotes the Cauchy principal value. Here we have employed
the relation
\begin{equation}
  \int_0^{\infty}e^{\pm iEt}\dd{t}=\pi\delta(E)\pm iP\frac{1}{E}.
\end{equation}
Then we have
\begin{equation}
  \int_0^{\infty}\cos\Delta tC_{\alpha}(t)\dd{t}=\frac{1}{2}[C_{\alpha}(\Delta)+C_{\alpha}(-\Delta)],\quad
  i\int_0^{\infty}\sin\Delta tC_{\alpha}(t)\dd{t}=\frac{1}{2}[C_{\alpha}(\Delta)-C_{\alpha}(-\Delta)].
\end{equation}
Therefore the dissipator can be written as
\begin{equation}
  \mathcal{D}_{\alpha}[\hat{\rho}_d(t)]=-[\hat{\sigma}_z,\hat{\varLambda}_{\alpha}\hat{\rho}_d(t)]+[\hat{\sigma}_z,\hat{\rho}_d(t)\hat{\varLambda}_{\alpha}^{\dag}],\quad
  \hat{\varLambda}_{\alpha}=\frac{1}{2}\mqty[C_{\alpha}(\Delta)+C_{\alpha}(-\Delta) & C_{\alpha}(\Delta)-C_{\alpha}(-\Delta)\\
  C_{\alpha}(-\Delta)-C_{\alpha}(\Delta) & -C_{\alpha}(\Delta)-C_{\alpha}(-\Delta)].
\end{equation}

\section{Convergence Analysis}
\label{app:convergence}

\begin{figure}[htbp]
\centerline{\includegraphics[]{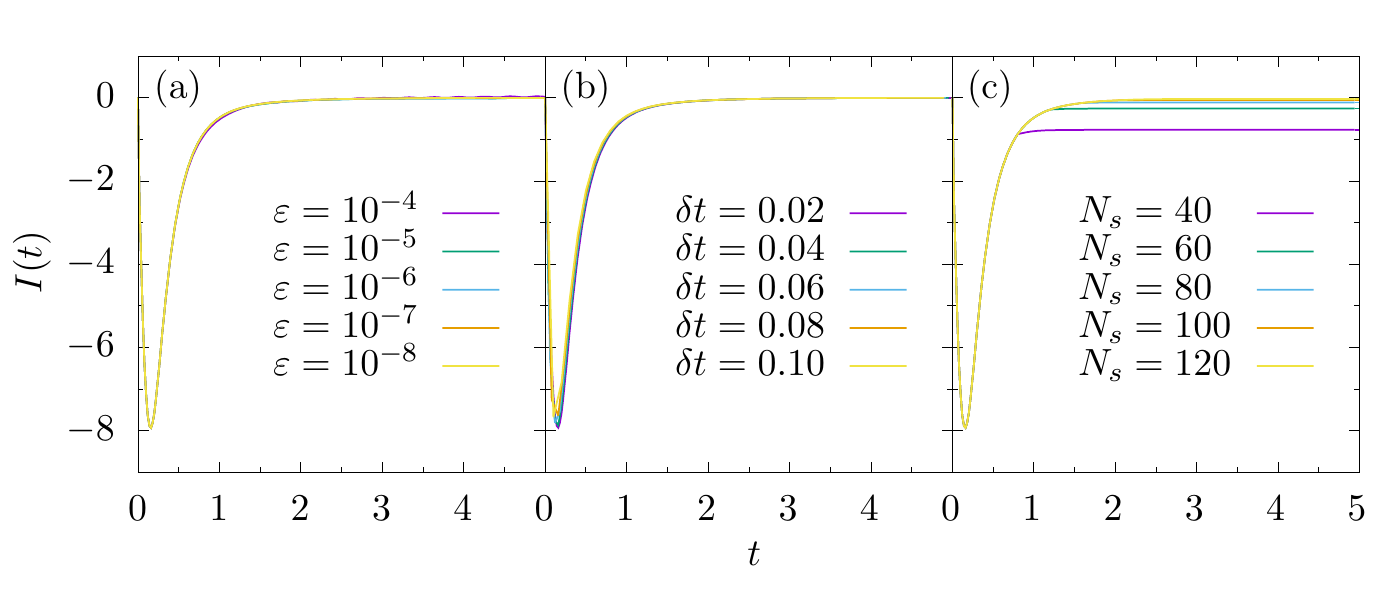}}
\caption{(a) The current $I(t)$ with $\delta t=0.02$, $N_s=200$ and
  different $\varepsilon$. (a) The current $I(t)$ with
  $\varepsilon=10^{-8}$, no $N_s$ and different time step $\delta
  t$. (b) The current with $\delta t=0.02$ and different $N_s$.}
\label{fig:d01}
\end{figure}

The convergence is controlled by parameters $\delta t$, $N_s$ and
$\varepsilon$. Here we use the single bath spin-boson model given in
Sec. \ref{sec:spin-boson-model-with-single-bath} to show the
convergence test. Here we demonstrate the case with a large coupling
strength $\lambda=0.5$ and the temperature is $T=1$. We first set a
small $\delta t=0.02$ and a fairly large $N_s=200$ to find a proper
SVD truncation parameter $\varepsilon$, the results are shown in
Fig. \ref{fig:d01}(a) where the current flows out from the bath $I(t)$
with different $\varepsilon$ are shown. The value
$\varepsilon=10^{-4}$ is too coarse that the curve is not even smooth,
and with $\varepsilon=10^{-6}$ the convergence is already archived. In
this article, we adopt a much finer value $\varepsilon=10^{-8}$ which
is also used in Ref. \cite{otterpohl2022-hidden}. In
Ref. \cite{fux2021-efficient}, a looser value $\varepsilon=10^{-6.5}$
is used.

In Fig. \ref{fig:d01}(b), the current with no $N_s$ and different time
step $\delta t$ are shown. It can be seen that the convergence can be
archived with fairly large $\delta t$. In order to show smooth details
at short time scale, we choose a smaller $\delta t=0.02$. In
Fig. \ref{fig:d01}(c), the current with $\delta t=0.02$ and different
$N_s$ are shown. It can be seen that small $N_s$ may cause
discontinuity of $I(t)$, see $N_s=40$ case. In this case, we have
$N_s\delta t=0.8$, therefore a discontinuity arises at
$t=0.8$. Similarly, when $N_s=60$ we have a discontinuity at
$t=1.2$. The presence of discontinuity is a sign that $N_s$ is not
large enough. When $N_s$ is large enough, the discontinuity would
disappear and we obtain convergent results. It can be seen that
$N_s=120$ is already large enough and in this article we adopt a even
larger value $N_s=200$. Basically, the model used in this article is
not difficult to converge, and for safety we have adopted stricter
parameters.

\providecommand{\newblock}{}

\end{document}